\documentstyle[12pt,moriond,epsf,psfig,colordvi]{article}
\newcommand{\alfs}{\mbox{$\alpha_s$}}

\newcommand{\mztwo}{\mbox{$M_Z^2$}}

\def\be{\begin{equation}}
\def\ee{\end{equation}}
\def\bea{\begin{eqnarray}}
\def\eea{\end{eqnarray}}
\def\beas{\begin{eqnarray*}}
\def\eeas{\end{eqnarray*}}
\def\np#1#2#3   {{\em Nucl. Phys.} {\bf#1}, #2 (#3) }
\def\pcps#1#2#3 {{\em Proc. Cam. Phil. Soc.} {\bf#1}, #2 (#3) }
\def\pl#1#2#3   {{\em Phys. Lett.} {\bf#1}, #2 (#3) }
\def\prep#1#2#3 {{\em Phys. Rep.} {\bf#1}, #2 (#3) }
\def\prev#1#2#3 {{\em Phys. Rev.} {\bf#1}, #2 (#3) }
\def\prl#1#2#3  {{\em Phys. Rev. Lett.} {\bf#1}, #2 (#3) }
\def\prs#1#2#3  {{\em Proc. Roy. Soc.} {\bf#1}, #2 (#3) }
\def\ptp#1#2#3  {{\em Prog. Th. Phys.} {\bf#1}, #2 (#3) }
\def\rmp#1#2#3  {{\em Rev. Mod. Phys.} {\bf#1}, #2 (#3) }
\def\rpp#1#2#3  {{\em Rep. Prog. Phys.} {\bf#1}, #2 (#3) }
\def\zp#1#2#3   {{\em Zeit. Phys.} {\bf#1}, #2 (#3) }
\def\epj#1#2#3   {{\em Eur. Phys. Jour.} {\bf#1}, #2 (#3) }
\begin{document}
\input psfig.tex
\newcommand{\linespace}[1]{\protect\renewcommand{\baselinestretch}{#1}
  \footnotesize\normalsize}
\begin{flushright} 
\typeout{need preprint number} 
UR-1519\\
CDF/PUB/ELECTROWEAK/PUBLIC/4596\\
May 21, 1998
\end{flushright}
\begin{center} 
\vspace*{1.5cm} 
{\large PARTON DISTRIBUTIONS, $d/u$, AND HIGHER TWISTS AT HIGH $X$}
\end{center} 
{\footnotesize
\begin{center}
\vspace{1.5cm}
\begin{sloppypar}
\noindent
 U.~K.~Yang \footnote{ To be published in proceedings of the 33rd Rencontres 
de Moriond: QCD and High Energy Hadronic Interaction, France, Mar. 1998.
Email: ukyang@fnal.gov}
A.~Bodek, and Q.~Fan
\vspace{15pt} 

 Department of Physics and Astronomy,
 University of Rochester, Rochester, NY 14627 \\             
\vspace{3pt}

\end{sloppypar}
\end{center}}

\begin{center} 
\vspace{3.5cm}

\end{center} 

\vspace{0.3in} 

\begin{abstract} 
We extract the ratio of the down ($d$) and up ($u$) parton distribution 
functions
(PDF's) from the ratio of deuteron and proton structure 
function $F_2^d/F_2^p$ measured by NMC.
We use corrections for nuclear binding effects 
in the deuteron, which are extracted 
from the  nuclear dependence of SLAC $F_2$ data.
Significant corrections to the $d$ quark distribution
in standard PDF's are required, especially at high $x$.
The corrected $d/u$ ratio is in agreement with the QCD prediction
of 0.2 at $x=1$.  The predictions for the $W$ asymmetry in 
hadron colliders using PDF's with the corrected $d/u$ ratio
are in much better agreement with recent CDF data at large rapidity.
Using the updated $d/u$ ratio and the most recent world
average for $\alpha_{s}$, we perform a NLO global fit to all DIS data
for $F_2$ and $R$, and estimate the size of the higher twist contributions
using both a renormalon model and an empirical model. 
We find that with the updated value of $\alpha_{s}$,
the magnitude of the higher twist terms is half the value 
of previous analysis.
With the inclusion of target mass and higher twist corrections, 
the standard NLO PDF's with the
updated $d/u$ ratio describe the SLAC $F_2$ data
up to $x=1.0$.  When the analysis is repeated in NNLO, 
we find that the additional 
NNLO contributions to $R$ account
for most of the higher twist effects extracted in the NLO fit.
The analysis in NNLO indicates that 
the higher twist effects in $R$, $F_2$ and $xF_3$ (e.g. GLS sum rule) are very small.

\end{abstract} 

\pagebreak

\section{Introduction}

 Recent work on parton distributions functions (PDF's)
in the nucleon has focussed
on probing the sea and gluon distribution at small $x$. The valence 
quarks distribution has been thought to be relatively well understood.
However, the precise knowledge of the $u$ and $d$ quark distribution
at high $x$ is very important at collider energies in searches for signals
for new physics at high $Q^2$.
In addition, the value of $d/u$ as $x \rightarrow 1$ is of theoretical
interest.
Recently, a proposed CTEQ toy model~\cite{toy} included
 the possibility of an additional contribution to the $u$ quark distribution
(beyond $x>0.75$) as an explanation for both the initial HERA high
 $Q^2$ anomaly~\cite{highQ2},
and for the jet excess at high-$P_t$  at CDF~\cite{CDFjet}. In this
communication we conclude that a re-analysis of data from
NMC and SLAC leads to a great improvement in the determination
of PDF's at large $x$.

\section{Extraction of $d/u$ at high $x$}

\begin{figure}
\centerline{\psfig{figure=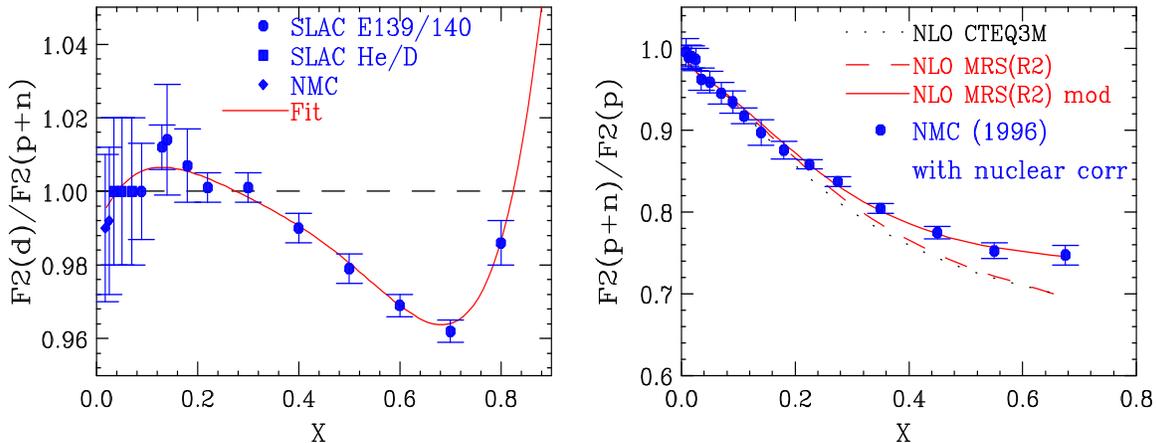,width=6.0in,height=2.3in}}
\caption{(Left) The total correction for nuclear 
effects (binding and Fermi motion) in the deuteron,
 $F_2^d/F_2^{n+p}$, as a function of $x$, extracted from fits to
the nuclear dependence of SLAC $F_2$ electron scattering
data. (Right) Comparison of NMC $F_2^{n+p}/F_2^p$ (corrected for nuclear
effects) and the prediction in NLO using the  MRS(R2) 
\leftline{PDF
with and without our proposed modification to the $d/u$ ratio.}}
\label{fig:f2dp}
\end{figure}

Information about valence quarks originates from the proton and neutron
structure function data. The $u$ valence quark distribution at high $x$
is relatively well
constrained by the proton structure function $F_2^p$.
However, the neutron structure function $F_2^n$, which is sensitive 
to the $d$ valence quark at high $x$,
is actually extracted from deuteron data.
 Therefore, there is an uncertainty in the $d$ valence quark 
distribution from the corrections for nuclear binding effects in the deuteron. 
In past extractions of $F_2^n$ from deuteron data,
only Fermi motion corrections
were considered, and other 
binding effects were assumed to be negligible.
Recently, the corrections for nuclear binding effects in the deuteron,
$F_2^d/F_2^{n+p}$, have been extracted empirically from
fits to the nuclear dependence
of electron scattering data from SLAC experiments
E139/140~\cite{GOMEZ}.
The empirical extraction uses a
model proposed by Frankfurt and Strikman~\cite{Frankfurt}, 
in which all binding effects 
in the deuteron and heavy nuclear targets
are assumed to scale with the nuclear density.
The correction extracted in this empirical way is also in agreement 
(for $x<0.75$) with recent purely theoretical
calculations~\cite{duSLAC} of nuclear binding effects in the deuteron.
The total correction for nuclear
binding in the deuteron is about 4\% at $x=0.7$
(shown in Fig. \ref{fig:f2dp}, left), and in a direction which is opposite
to what is expected from the previous models which only included
the Fermi motion effects.

 The ratio $F_2^d/F_2^p$ is directly related to $d/u$. In leading order QCD,
 $2F_2^d/F_2^p -1 \simeq (1+4d/u)/(4+d/u)$ at high $x$.
  We perform a NLO analysis on the precise
  NMC $F_2^d/F_2^p$ data~\cite{NMCf2dp} to extract $d/u$
  as a function of $x$.
 We extract the ratio $F_2^{p+n}/F_2^p$
 by applying the nuclear binding correction
 $F_2^d/F_2^{n+p}$  to the $F_2^d/F_2^p$ data.

\begin{figure}
\centerline{\psfig{figure=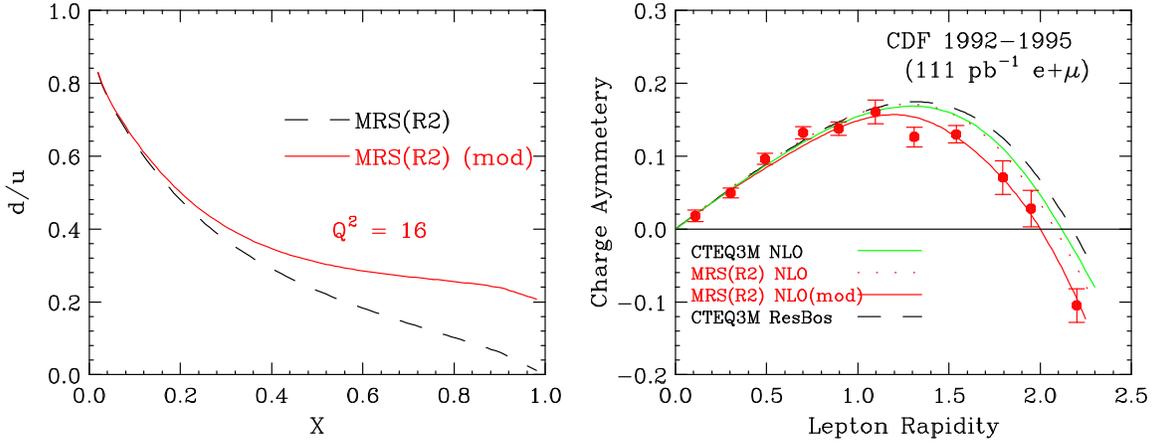,width=6.0in,height=2.3in}}
\caption{(Left) The $d/u$ distribution at $Q^2$=$16$ as a function of $x$. 
The standard MRS(R2) is compared to our modified MRS(R2). 
(Left) Comparison of the CDF $W$ asymmetry data with NLO standard
CTEQ3M, MRS(R2), and modified MRS(R2) as a function of the lepton rapidity.
The standard CTEQ3M with 
\leftline{a resummation calculation is also shown 
for comparison.}}
\label{fig:dou}
\end{figure}

 As shown in Fig. \ref{fig:f2dp}(right), the standard PDF's do
 not describe the extracted $F_2^{p+n}/F_2^p$.
 Because the $u$ distribution is relatively well constrained,
 we find a correction term to  $d/u$ in the standard PDF's,
 as a function of $x$ by only varying the $d$ distribution to fit the data.
 The correction term is  parametrized  
 as a simple quadratic form, $\delta (d/u) = (0.1\pm0.01)(x+1)x$
 for the MRS(R2) PDF,
 where the corrected $d/u$ ratio $(d/u)'$,
 is $(d/u)' = (d/u) + \delta (d/u)$.
 Based on this correction,
 we obtain a  MRS(R2)-modified PDF as shown in Fig \ref{fig:dou}.
 The corrections to other PDF's such as CTEQ3M is similar.
 The NMC data, when corrected for nuclear binding effects
in the deuteron, clearly indicate that $d/u$ in the
 standard PDF's~\cite{MRSR2,CTEQ3M} is significantly underestimated 
 at high $x$ as shown in Fig. \ref{fig:dou}.
 Fig.\ref{fig:dou} (left) also shows that the modified $d/u$ ratio 
 approaches
 $0.2\pm0.02$ as $x \rightarrow 1$, in agreement with a QCD
 prediction~\cite{Farrar}. In contract, if the
deuteron data is only corrected for Fermi motion effects (as
was mistakenly done in the past) both the $d/u$ from data and the $d/u$
in the standard PDF's fits
approach $0$ as $x \rightarrow 1$.

Information (which is not affected by the corrections
for nuclear effects in the deuteron)
on $d/u$ can be extracted from $W$ production data in hadron
colliders.
Fig.\ref{fig:dou} (right) shows that the predicted $W$ asymmetry calculated
with the DYRAD NLO QCD program
using our modified PDF is
in much better agreement with recent CDF data~\cite{Wasym} at large
rapidity than standard PDF's.

When the modified PDF at $Q^2$=$16$ is evolved to $Q^2$=$10,000$ 
using the DGLAP NLO equations, we find that
the modified $d$ distribution at $x=0.5$ is increased by about 40 \% 
in comparison to the standard $d$ distribution.
The modified PDF could have a significant impact
on the prediction of the cross sections in the HERA high $Q^2$ region because 
the charged current scattering with positrons is on $d$ quarks only.
It also impacts neutral current scattering because of the large
coupling of the $Z$ to $d$ quarks at high $Q^2$.

\section{Higher twists effects at high $x$}

Since all the standard PDF's, including our modified version are
fit to data with $x$ less than 0.75, we now
investigate the validity of the modified MRS(R2) at very high
$x$  by comparing to  $F_2^p$ data at SLAC.
Although the SLAC data at very high $x$ are at reasonable values
of $Q^2$ $(7<Q^2<31 ~GeV^2)$, there are in a region in which
 non-perturbative effects such as target mass
 and higher twist are very large.
We use the Georgi-Politzer calculation~\cite{GPtm}
for the target mass corrections (TM). These involve
using the scaling variable 
$\xi=2x/(1+\sqrt{1+4M^2x^2/Q^2})$ instead of $x$.
Since a complete calculation of higher twist
effects is not available, the very low $Q^2$ data is
used to obtain information on the size of these terms.

We use two approaches in our investigation of the higher
twist contributions, 
an empirical method, and the renormalon model. 
In the empirical approach, the higher twist contribution is evaluated
by adding a term $h(x)/Q^2$ to the perturbative QCD (pQCD) prediction 
of the structure function (including target mass effects).
 The $x$ dependence of the higher
twist coefficients $h(x)$ is fitted to the global DIS $F_2$ 
(SLAC, BCDMS, and NMC) data~\cite{SLACF2,BCDMSF2,NMCF2} in the kinematic region 
($0.1<x<0.75$, $1.25<Q^2<260 ~GeV^2$)
with the following form, 
$F_2 = F_2^{pQCD+TM}(1+h(x)/Q^2)f(x)$. 
Here $f(x)$ is a floating factor to investigate possible
$x$ dependent corrections to our modified PDF.
A functional form, $a(\frac{x^b}{1-x}-c)$ 
for $h(x)$ is used in the higher twist fit to
estimate the size of the higher twist terms above $x=0.75$.
The SLAC and BCDMS data are normalized to the NMC data.
In the case of the BCDMS data, a systematic shift error $\lambda$ is allowed
to account for the point-to-point correlated systematic errors.
The empirical higher twist fits with the modified NLO MRS(R2) pQCD
prediction with TM
have been performed simultaneously
on the proton and deuteron $F_2$ data with 11 free parameters
(2 relative normalizations and 3 parameters for $C(x)$ per target
and the BCDMS $\lambda$).
We find that empirical higher twist fit describes the data well
($\chi^2/DOF=843/805$). The size of the higher twist contributions
in the  proton and deuteron are similar.
The magnitude is  almost half of those extracted
in previous analysis of SLAC/BCDMS
data~\cite{Virchaux}.
This is because that analysis was based on \alfs(\mztwo) $=0.113$,
while the MRS(R2) PDF uses \alfs(\mztwo) $=0.120$,
which is close to the current world average.


\begin{figure}[ht]
\centerline{\psfig{figure=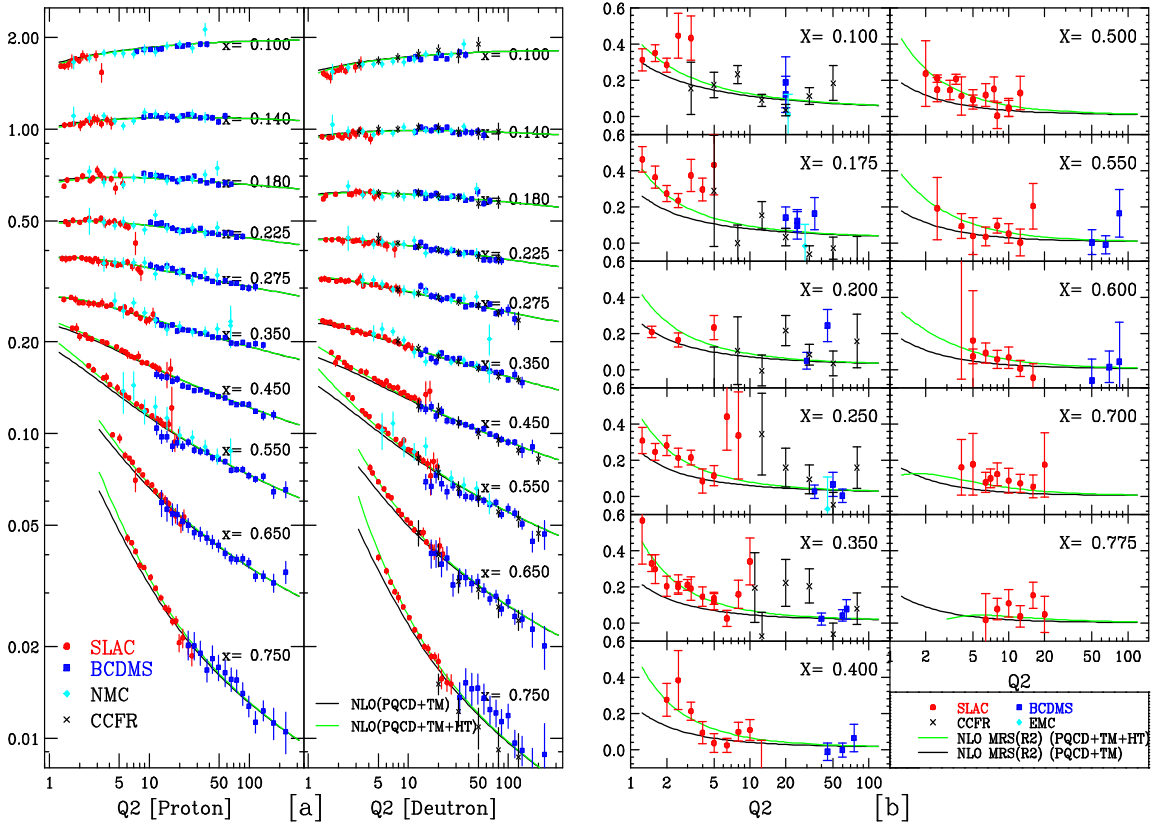,width=6.0in,height=4.3in}}
\caption{The description of higher twist fit using the renormalon model
with the modified NLO MRS(R2) PDF. The CCFR neutrino data is also shown 
for comparison.
(Left) Comparison of $F_2$ and NLO prediction
with and without higher twist contributions. 
(Right) Comparison of $R$ and NLO prediction
with and without 
\leftline{the renormalon higher twist contributions.}}
\label{fig:disht}
\end{figure}

In the renormalon model approach~\cite{renormalon}, the 
model predicts the complete $x$ dependence of the higher twist
contributions 
 to  $F_2$, $2xF_1$, and $xF_3$ ,with only two unknown parameters
$A_2$ and $A_4$. 
We extract the $A_2$ and $A_4$ parameters, which
determine the overall level of the  $1/Q^2$
and $1/Q^4$ terms by fitting to the global
data set for $F_2$ and $R (= F_2(1+4Mx^2/Q^2)/2xF_1 - 1)$.
The values of $A_2$ and $A_4$ for the proton and deuteron
are same in this model.
The $x$ dependence of $2xF_1$ differs from that of $F_2$ but is same as that
of $xF_3$ within a power correction of $1/Q^2$. 
Our fits  can also be used to estimate the size of the higher twist effects
in $xF_3$ (e.g. the GLS sum rule).
The higher twist fit in this approach has employed the same procedure 
as the empirical method.  Fig. \ref{fig:disht} shows that the model yields
description of $x$ dependence of higher twist terms in
 both $F_2$ and $R$
with just the two free parameters ($\chi^2/DOF=1577/1045$).
The CCFR neutrino data~\cite{CCFRF2}
is shown for comparison though it is not used in the fit.
The extracted values of $A_2$ are $-0.093 \pm 0.005$
 and $-0.101 \pm 0.005$, for proton and deuteron,
 respectively. The contribution of $A_4$ is found  to be negligible. 
We find that the floating factor $f(x)$ for the deuteron deviates from 1
and is also bigger than
that for the proton, unless the modified MRS(R2) PDF is used.
This reflects our earlier conclusion
that the standard $d$ distribution is underestimated at high $x$ region.
As expected, the extracted $A_2$ value
is half of the previous estimated value~\cite{renormalon} 
of $A_2$ based on SLAC/BCDMS analysis. 
Since both of these approaches yield a reasonable description for the higher
twist effects, we proceed to compare the predictions of the modified
PDF's (including target mass and renormalon higher twist corrections) to the
SLAC proton $F_2$ data at very high $x$ ($0.7<x<1$). 

\section{Parton distributions functions at very high $x$}

\begin{figure}
\centerline{\psfig{figure=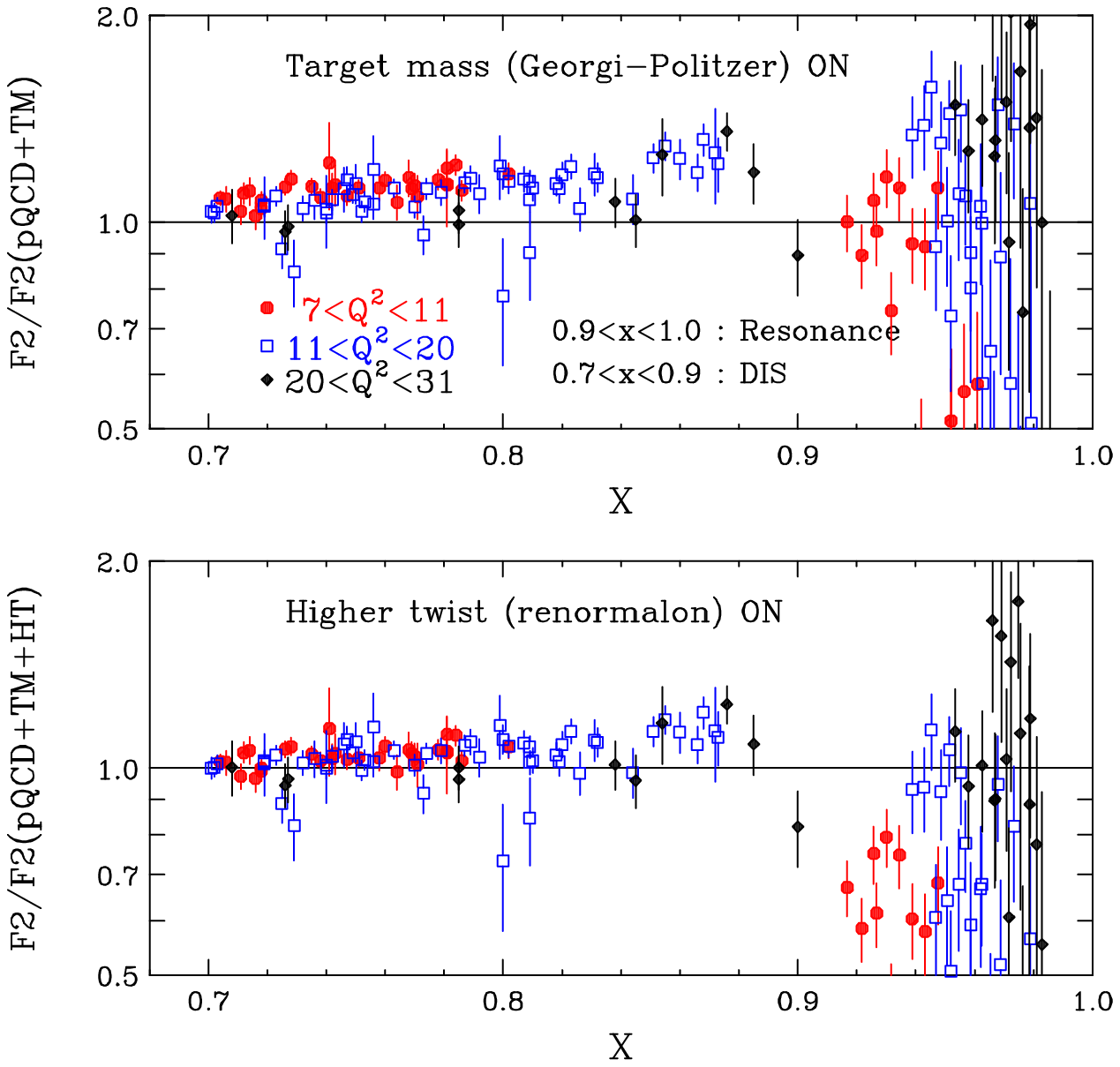,width=4.2in,height=3.0in}}
\caption{The ratio of SLAC $F_2^p$ data at high $x$
to the predictions of the modified MRS(R2).
(Top) The data are compared to the NLO pQCD prediction
with the inclusion of target mass effects.
(Bottom) The data are compared to the NLO pQCD prediction 
with the inclusion of
target mass and renormalon higher twist effects. 
}
\label{fig:highx}
\end{figure}

There is a wealth of SLAC data~\cite{SLACres} in the region up to $x=0.98$ 
and intermediate $Q^2$ $(7<Q^2<31 ~GeV^2)$. 
Previous PDF fits have not used these data. 
We use the estimate of the higher twist effects 
from the models, based on the data (below $x<0.75$) described above.
Note that the data for $x>0.75$ is in the DIS region,
and the data for $x>0.9$ is the resonance region.
It is worthwhile to investigate the resonance region also because
from duality arguments~\cite{Bloom} it is expected  
that the average behaviour of the resonances and elastic peak
should follows the DIS scaling limit curve.

Fig.\ref{fig:highx} shows the ratio of the SLAC data at high $x$
to the predictions of the modified MRS(R2).  
With the inclusion of target mass and the renormalon higher twist
effects, the very high $x$ data from SLAC is remarkably well described 
by the modified MRS(R2) up to $x=0.98$. The data is somewhat lower than
the fit  at $x>0.9$ and $7<Q^2<11 ~GeV^2 $.
This could be due to the missing elastic
contribution (as expected from duality arguments). The good decription
of the data by the modified MRS(R2) is also achieved 
using the empirical estimate ($h(x)/Q^2$) of higher twist effects.
From these comparisons,
we find that the SLAC $F_2$ data do not favor the CTEQ Toy model
which proposed an additional $u$ quark contribution
at high $x$ (beyond $x>0.75$) 
as an explanation of the initial HERA high $Q^2$ anomaly
and the CDF high-$P_t$ jet excess.
This model (with an additional 0.5\%
component of $u$ quarks) overstimates the SLAC data 
by a factor of two at $x = 0.9$ (DIS region).

\section{Conclusion}

We find that nuclear binding effects in the deuteron
play a significant role in understanding of the $d/u$
at high $x$ region. The modified PDF's with our $d/u$ correction
are in good agreement with the prediction of QCD at $x=1$, and with the
CDF $W$ asymmetry data. With the inclusion of target mass and 
higher twist corrections, the modified PDF's describe all DIS data
not only up to the very highest $x$, but also all
the way down to $Q^2 = 1 ~GeV^2$.

Note that when the analysis is repeated in NNLO~\cite{NNLO}, 
We find that the additional 
NNLO contributions to $R$ account
for most of the higher twist effects extracted in the NLO fit.
The analysis in NNLO indicates that the highest twist corrections
to the GLS sum rule are very small (the fractional contribution
to the pQCD GLS sum rule is $-0.0058/Q^2-0.013/Q^4$).


\vspace{-.15in} 
\section*{References}
\begin{thebibliography}{99}
\small

\bibitem{toy}
S. Kuhlmann, H.L. Lai, W.K. Tung, \pl{B409}{271}{1997}

\bibitem{highQ2}
C. Adloff {\em et al}., \zp{C74}{191}{1997} ,
J. Breitweg {\em et al}.,\zp{C74}{207}{1997} 


\bibitem{CDFjet}
F. Abe {\em et al}.,  \prl{77}{438}{1996} 


\bibitem{GOMEZ}
J. Gomez {\em et al}., \prev{D49}{4348}{1994} 

\bibitem{Frankfurt}
Frankfurt and Strikman, \prep{160}{235}{1998}

\bibitem{duSLAC}
 W. Melnitchouk and A.W. Thomas, \pl{B377}{11}{1996}

 W. Melnitchouk and J.C. Peng,  \pl{B400}{220}{1997}

\bibitem{NMCf2dp}
 M. Arneodo {\em et al}., \np{B487}{3}{1997} 

%
\bibitem{MRSR2}
A.D. Martin {\em et al}.,\pl{B387}{419}{1996} 

\bibitem{CTEQ3M}
H.L. Lai {\em et al}., \prev{D51}{4723}{1995} 

\bibitem{Farrar}
G.R. Farrar and D.R. Jackson, \prl{35}{1416}{1975} 


\bibitem{Wasym}
A. Bodek, in proceedings of EW Moriond (1998).


\bibitem{GPtm}
H. Gerogi and H.D. Politzer, \prev{D14}{1829}{1976}

\bibitem{SLACF2}
L.W. Whitlow  {\em et al}., \pl{B282}{475}{1992} 
 
\bibitem{BCDMSF2}
A.C. Benvenuti {\em et al}., \pl{B223}{485}{1989}

A.C. Benvenuti {\em et al}., \pl{B237}{592}{1990} 

\bibitem{NMCF2}
M. Arneodo {\em et al}., \np{B483}{1997}{3}{1997} 
\bibitem{Virchaux}
M. Virchaux and A. Milsztajn, \pl{B274}{221}{1992} 


\bibitem{CCFRF2}
W.G. Seligman {\em et al}., \prl{79}{1213}{1997} 

\bibitem{renormalon}
M. Dasgupta and B.R. Webber, \pl{B382}{273}{1996} 


\bibitem{SLACres}
P. Bosted  {\em et al}., \prev{D49}{3091}{1994} 

\bibitem{Bloom}
E.D. Bloom and F.J. Gilman  \prl{25}{1140}{1970} 

\bibitem{NNLO}
E. Zijlstra and W.L. van Neerven \np{B383}{525}{1992}
\end{thebibliography}
\end{document}